# Beyond the Isotropic Lifshitz Endpoint

Tom T. S. Chang

Massachusetts Institute of Technology

Cambridge, MA 02139 USA

Email: tom.tschang@gmail.com

**Abstract.**

   The puzzle of the disappearance of isotropic Lifshitz points in condensed matter physics is explained from the point of view of the Wilsonian renormalization group. In analogy to the commensurate ideas of metamagnetic phase transitions, we describe the physics of thermodynamic states beyond an isotropic Lifshitz endpoint. Such phenomenon may be understood in terms of a statistically isotropic environment of coexisting multi-incommensurate helicoidal states. In addition to the magnetic and condensed matter discussions, we consider also an interesting example in the context of dynamical evolution.

## I. Introduction

   Helical ordering in magnetic systems was initially discovered by Yoshimori [1], Kaplan [2], and Villain [3] in terms of the mean field theory. The tricritical or Lifshitz behavior of the helical, ferromagnetic, and disordered phases was first examined in terms of the renormalization group (RG) by Hornreich et al. [4] by considering the non-quadratic effect of the critical propagator of the effective action for such a system. Later, this idea was generalized by Nicoll et al. [5] to arbitrary order and also with anisotropy [6]. In these studies, it was discovered that for isotropic systems, critical points may not exist beyond certain value of the isotropic Lifshitz order $L$ which needs not to be an integer. In this letter, we study in detail when and why the non-existence of this criticality set in as well as the interesting thermodynamic behavior beyond this isotropic Lifshitz endpoint.

## II. Renormalization-Group Analysis

Exact RG formulations with UV cutoffs at momentum scales much smaller than Planckian [7-9] are sufficient for our discussion. We search for infrared fixed points (IRFPs) from such a formulation perturbatively using the method of Gaussian eigenfunction expansions as first described by Nicoll et al. [10] and Chang et al. [11] and extended later to systems with generalized critical propagators [5,6] ($k^\sigma$, $\sigma = 2L$ for isotropic and $\sum k_i^{\sigma_i}$ for isotropic systems). Briefly, the method first determines the Gaussian eigenvalues and eigenfunctions of the RG equation and then search for small nontrivial IRFPs by assuming one of the Gaussian eigenvalues and the related nontrivial fixed point effective action to be small. The procedure for perturbative calculations to first order may be simplified by assuming the anomalous dimension $\eta$ for these calculations to be of higher order (which may be verified a posteriori). The RG equation then reduces to a nonlinear partial differential equation (NPDE).

For example, for an isotropic generalized critical propagator and the isotropic ordering of an n-component vector field $\mathbf{s}(\mathbf{x})$, the Gaussian eigenfunctions for the NPDE are generalized Laguerre polynomials $Q_p = L_p^{n/2-1}\left((d-\sigma)/4)s^2\right)$, and the corresponding eigenvalues are $\lambda_p = d + (\sigma - d)p$, where $\sigma$ is the exponent of the critical propagator. We now define an expansion parameter $\varepsilon_\vartheta(\sigma) = d + \vartheta(\sigma - d)$ and assume it to be small, where $\vartheta$ is the order (or multiplicity) of the phase transition. The search for a nontrivial small fixed point effective action $A^* = c\varepsilon_\vartheta(\sigma)Q_\vartheta$ may be accomplished by expanding the effective action in the NPDE generator in terms of the Gaussian eigenfunctions and equating the terms to order $\varepsilon_\vartheta^2(\sigma)$ to evaluate $c$. Linearizing about this small but nontrivial fixed point, we can calculate the perturbed eigenvalues $\overline{\lambda}_p$ to first order in $\varepsilon_\vartheta(\sigma)$. The results are:

$$1 = -c\langle \vartheta, \vartheta; \vartheta \rangle / \langle \vartheta; \vartheta \rangle, \text{ and} \tag{1a}$$

$$\overline{\lambda}_p = \lambda_p - 2\varepsilon_\vartheta(\sigma)\langle \vartheta, p; p \rangle / \langle \vartheta, \vartheta; \vartheta \rangle \tag{1b}$$

where $\langle i, j; k \rangle$ is the projection onto $Q_k$ of the quadratic part of the RG equation applied to $Q_i$ and $Q_j$, $\langle k; k \rangle$ is the normalization factor, and $\langle \vartheta, p; p \rangle$ may be expressed in closed form in terms of the binomial coefficients as follows:

$$\langle \vartheta, p; p \rangle = \sum_{j=0}^{[\vartheta/2]} \binom{p}{j}\binom{p-1+n/2}{j}\binom{2p-2j}{\vartheta-2j} \tag{1c}$$

with [•] being the integer part of the enclosed quantity.

These results lead to physically realizable attractive IRFPs when the spatial dimension lies within the upper $d_+$ (above which mean field theory applies) and lower $d_-$ (below which catastrophic divergences occur) borderline dimensions given below [12]:

$$d_+ = \sigma\vartheta/(\vartheta-1) \quad \text{and} \quad d_- = \sigma \tag{2}$$

Thus, for realizable isotropic IRFPs for $d = 3$, we must have $\sigma \leq 3$.

For anisotropic generalized Lifshitz ordering, the physically realizable dimension lies within the upper and lower dimensions that satisfy [12]

$$\sum d_i / \sigma_i = \vartheta/(\vartheta-1) \tag{4a}$$

$$\sum d_i / \sigma_i = 1 \tag{4b}$$

respectively. Unlike the isotropic situation, anisotropic generalized Lifshitz ordering of the above type are generally realizable for $d = 3$. For example, the upper and lower borderline dimensions for a uniaxial generalized critical propagator with only one of the 3-dimensional wave vector $k_1$ raised to a power larger than 2 are:

$$d_+(\vartheta, \sigma_1) = (3\vartheta - 1)/(\vartheta - 1) - 2/\sigma_1 \tag{5a}$$

$$d_-(\vartheta, \sigma_1) = 3 - 2/\sigma_1 \tag{5b}$$

Thus, $d_- < 3 \leq d_+$ for $\vartheta \leq \sigma_1 + 1$.

### III. Beyond the Isotropic Lifshitz Endpoint

To understand the thermodynamics beyond the physically realizable isotropic Lifshitz state, let us consider first a somewhat similar phenomenon: the metamagnetic phase transition in condensed matter physics. Generally for an antiferromagnet, its critical (Néel) point decreases with the application of an applied magnetic field. At some critical field, however, the criticality for a metamagnet terminates at a tricritical point [13]. Beyond this field, the associated symmetry breaking leads to the first order transition to the tri-coexistence of two opposite antiferromagnetic-sublattice states and the paramagnetic state. In other words, it is the critical point of the tricoexistence line.

Returning to the discussion of the appearance of the isotropic Lifshitz endpoint, the physics for the point of demarcation along the isotropy line in the Affine space of critical propagator exponents (Fig. 1) may therefore be interpreted as the critical point of the statistically isotropic coexistence curve of multiphases of anisotropic Lifshitz states. Any thermodynamic state along the isotropic multiphase line beyond this endpoint is characterized by the statistically isotropic coexistence of subdomains of regions spanned by the neighborhoods (phases) of the various admissible anisotropic IRFPs and the disordered state. Such a coexistence state is determined by the statistically averaged propagator singularities of the IRFPs.

**IV. An Example**

We consider the example of gravitational evolution at large cosmological spatial (or small *k*) scales due to classical fluctuations. Our attention will be restricted to the matter-dominated era of non-relativistic dynamic motion. We therefore assume: (i) the gravitational field is weak, (ii) the variations of fields are slow in time, and (iii) the particles are nonrelativistic. Thus, the Einstein-Hilbert effective action reduces to:

$$S \cong -\int (8\pi G)^{-1}[(\nabla \varphi)^2 + \lambda)]d^3xdt - m\int \varphi dt + \sum (m/2)\int v^2 dt \qquad (1)$$

where $\varphi = h_{00}/2$ is the gravitational potential, $g^{00} \approx 1 - h_{00}$ with ($h_{00} \ll 1$) is the only surviving metric element, $G$ is the gravitational constant, $\lambda$ is the cosmological constant and $(m,v)$ are the masses and corresponding velocities.

Existing considerations [14] of IRFPs in gravitational evolution require the scale running of the gravitational constant to retain its natural dimension as the fixed point is approached. Here, we shall postulate that $G$, the propagator coupling constant responsible for the ordering of the gravitational potential $\varphi$ in the effective action, behaves as most diagonalized component of coupling constants do in acquiring power-law singular characteristics as $k \to 0$ (e.g., as $k^{-\alpha}$ for isotropic propagation where $\alpha$ is the singularity index). Thus, for isotropic evolution, the critical propagator for (1) will be assumed to scale as $k^\sigma$ at criticality, where $\sigma = 2 + \alpha$. For anisotropic propagation, we shall assume the critical propagator to behave as $\sum |\mathbf{k}_i|^{\sigma_i}$, where each $\mathbf{k}_i$ is a $d_i-$ dimensional vector and $\sigma_i = 2 + \alpha_i$. The generalized critical propagator (or generalized Lifshitz) exponents, $\sigma$ and $\sigma_i$, will be considered real and positive but not necessarily integers [5,6]. For our interests here, we shall consider the situations for physically realizable gravitational dynamics at $d = 3$ dimensions.

At small spatial scales (such as stellar distances), we do not expect much appreciable scale-running effects of $G$ (and $\lambda$). Beyond the intermediate (larger than galactic) scales, we expect the scale-running effects of $G$ and $\lambda$ begin to come into play. Initially, the evolution at such scales can be anisotropic and this may be one of the reasons for the continuing development of large-scale structures in the Universe.

Although the dynamic interactions during the cosmological evolution will include other orderings in addition to the gravitational potential (such as those related to the fields $m$ and $\mathbf{v}$ in (1)), we do not expect these effects to influence the helicoidal orderings due to the scale-running of the propagator coupling constant $G$. In particular, we notice that the lower borderline dimension generally depends on the critical propagator exponents only. We may therefore use the RG results of Eqs. (2) and (4b) for $d_-$ to understand the gravitational phenomenon by setting $n = 1$, with the scalar field $s$ representing the gravitational potential $\varphi(x)$. Such complex effects (CILOMAS) involving helicoidal orderings have been treated in detail by Chang [15] in a Physics Letter.

Beyond the intermediate scales, we expect the evolutional dynamics to become more and more isotropic. When full isotropy is statistically achieved at very large spatial scales, we may discuss the gravitational evolution in terms of the Friedmann-Robertson-Walker (FRW) paradigm. For a late matter-dominated ($p = 0$) flat Universe, the Friedmann equation and the conservation equation of the energy-momentum tensor are:

$$(\dot{a}/a)^2 = (8\pi/3)G\rho_m + \lambda/3 \tag{6a}$$

$$\dot{\rho}_m + 3(\dot{a}/a)\rho_m = 0 \tag{6b}$$

where $a(t)$ is the scale factor of the expanding Universe, $\rho_m(t)$ is the matter density, and the over dot indicates differentiation with respect to time $t$. We shall follow the arguments given by Bonanno and Reuter [14] and relate the momentum and time in leading order in the infrared as $k = \xi/t$ where $\xi > 0$ is a constant. Thus $G(k = \xi/t) \equiv G(t) \sim t^\alpha$ at large spatial scales and the scale running of the cosmological constant $\lambda(k = \xi/t) \equiv \lambda(t)$. We close the above equations with the Bianchi identity as the consistency condition:

$$\dot{\lambda} + 8\pi\rho_m \dot{G} = 0 \tag{6c}$$

It was shown by Kalligas et al. [16] that Eqs. (6) may be integrated exactly for power law variations of $G = G_0 t^\alpha$ for $\alpha \neq -2$. The results are

$$\rho_m(t) = [(2+\alpha)/(12\pi G_0)]t^{-(2+\alpha)} \tag{7a}$$

$$\lambda(t) = [\alpha(2+\alpha)/3]t^{-2} \tag{7b}$$

$$a(t) = const \times t^{(2+\alpha)/3} \tag{7c}$$

We may then estimate the value of the singularity index $\alpha$ from (7) based on observational data. For example, we can estimate its value from the density parameter $\Omega_m = \rho_m/(\rho_m + \rho_\lambda)$ with the vacuum energy density, $\rho_\lambda \equiv \lambda/(8\pi G)$, using (7a,b). Taking the value of $\Omega_m \approx 0.321$ from the July 2018 Planck result [17], we find $\alpha \approx 4.230$. This in turn gives $\Omega_\lambda = \rho_\lambda/(\rho_m + \rho_\lambda) \approx 0.679$ and the deceleration parameter $q = -a\ddot{a}/\dot{a}^2 \approx -0.519$. Both values are very close to the presently accepted observational values and thus providing the natural explanation of the present day accelerating expansion of the Universe [18,19].

We know that for physically realizable isotropic generalized Lifshitz IRFPs we must have $\sigma \leq 3$ (or $\alpha \leq 1$). Thus for $d = 3$, beyond the critical end state at $\sigma = 3$ (or $\alpha = 1$), any statistically isotropic dynamic state will be that of a multi-phase coexistence state of subdomains of intermediate cosmic scales with various physically realizable helical and nonhelical structures (Fig. 1).

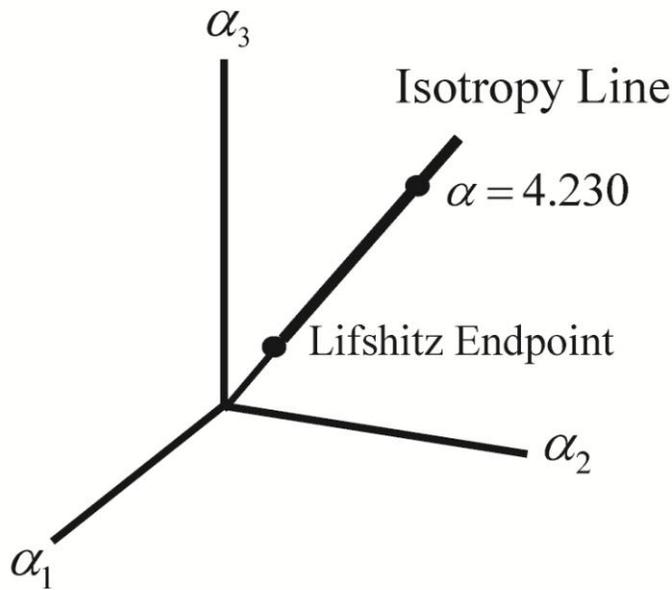

**Fig. 1.** Affine space of singularity indices $\alpha_i \equiv \sigma_i - 2$ for $d = 3$. Thin line segment: range of realizable isotropic fixed points. Thick line: domain of coexisting multi-phases.

**V. Summary**

We considered the thermodynamics beyond the admissible range of isotropic Lifshitz fixed points in RG calculations. Such phenomenon is akin to the well-known result of tricritical-tricoexistence effect of metamagnetic phase transitions in condensed matter physics. The isotropic Lifshitz endpoint is the critical point of the statistically isotropic multiphase coexisting thermodynamic state of subdomains of anisotropic helicoidal neighborhoods. An example of cosmic evolution based on the effective action with scale-running gravitational and cosmological constants at large spatial scales due to classical fluctuations is considered. The gravitational symmetry-breaking effects at intermediate (e.g., galaxy) scales are related to the development and formation of cosmic structures with multifractal characteristics [15,20,21]. At large cosmological scales, the FRW formulism leads to the understanding of statistically isotropic multi-phase coexisting states. We presently live in a matter-dominated statistically isotropic fractal Universe with a gravitational singularity index of $\alpha \approx 4.230$. The result provides a natural explanation to the cosmic accelerated expansion.

ORCID iD  https://orcid.org/0000-0003-3553-7572